\documentclass[5p]{elsarticle}

\usepackage{color}
\usepackage{graphicx}
\usepackage{float}
\usepackage{amsmath}
\usepackage{hyperref}

\newcommand{\macro}[1]{\textcolor{black}{#1}} 
\newcommand{\ldvw}{\macro{LigoDV-web}}
\newcommand{\ldv}{\macro{LigoDV}}

\newcommand{\ldvwusers}{\macro{\ensuremath{634}}}
\newcommand{\ldvwsessions}{\macro{\ensuremath{33,861}}}
\newcommand{\ldvwplots}{\macro{\ensuremath{139,875}}} 

\renewcommand{\today}{\number\day\space\ifcase\month\or
  January\or February\or March\or April\or May\or June\or
  July\or August\or September\or October\or November\or December\fi
  \space\number\year}

\begin{document}

\title{\ldvw{}: Providing easy, secure and universal access to a large distributed scientific data store for the LIGO Scientific Collaboration}


\address[csuf]{Gravitational-Wave Physics and Astronomy Center and Department of Physics, California State University Fullerton, Fullerton, CA 92831, USA}
\address[aei]{Albert-Einstein-Institut, Max-Planck-Institut f\"ur Gravi\-ta\-tions\-physik, D-30167 Hannover, Germany}
\address[cit]{LIGO, California Institute of Technology, Pasadena, CA 91125, USA}
\address[lsu]{Louisiana State University, Baton Rouge, LA 70803, USA}

\author[csuf]{J.~S.~Areeda}
\author[csuf]{J.~R.~Smith}
\author[aei]{A.~P.~Lundgren}
\author[cit]{E.~Maros}
\author[lsu]{D.~M.~Macleod}
\author[cit]{J.~Zweizig}


\begin{abstract}
Gravitational-wave observatories around the world, including the Laser Interferometer Gravitational-wave Observatory (LIGO), record a large volume of gravitational-wave output data and auxiliary data about the instruments and their environments. These data are stored at the observatory sites and distributed to computing clusters for data analysis.  \ldvw\ is a web-based data viewer that provides access to data recorded at the LIGO Hanford, LIGO Livingston and GEO600 observatories, and the 40m prototype interferometer at Caltech. The challenge addressed by this project is to provide meaningful visualizations of small data sets to anyone in the collaboration in a fast, secure and reliable manner with minimal software, hardware and training required of the end users. \ldvw\ is implemented as a Java Enterprise Application, with Shibboleth Single Sign On for authentication and authorization and a proprietary network protocol used for data access on the back end. Collaboration members with proper credentials can request data be displayed in any of several general formats from any Internet appliance that supports a modern browser with Javascript and minimal HTML5 support, including personal computers, smartphones, and tablets. To date \ldvwusers\ unique users have visited the \ldvw\ website in a total of \ldvwsessions\ sessions and generated a total of \ldvwplots\ plots. This infrastructure has been helpful in many analyses within the collaboration including follow-up of the data surrounding the first gravitational-wave events observed by LIGO in 2015. 

\end{abstract}

\maketitle

\section{Introduction}
\label{Section:Intro}
The Advanced Laser Interferometer Gravitational-Wave Observatory (Advanced LIGO)~\cite{aligo} is a project to directly measure gravitational waves, ripples in space and time that were predicted by Einstein's General Relativity, from astronomical sources. In 2015, the project observed two black-hole merger events~\cite{gw150914,gw151226,O1bbh} inagurating the field of gravitational-wave astronomy. The science related to LIGO is carried out by the  LIGO Scientific Collaboration, with more than 1,000 members worldwide. Along with the two Advanced LIGO observatories in Hanford, WA, and Livingston, LA, there are also single observatories in Italy (Advanced Virgo)~\cite{avirgo}, Japan (Kagra)~\cite{kagra}, and Germany (GEO600)~\cite{geo}. In addition, prototype interferometers such as the Caltech 40m lab~\cite{40m} are testing new technologies. Together, these instruments produce a vast store of scientific data. 
Other major astronomy and physics projects with complex and large data sets (and often unique data formats) have implemented custom web-based solutions to share and plot their data, e.g.~\cite{DCSviewer,exoarchive,mast}. 
In this paper we describe a software suite, \ldvw, that has been developed to provide secure and universal data access to members of the LIGO Scientific Collaboration.  

\subsection{Data sources and characteristics}

\begin{figure*}
  \centering
  \includegraphics[width=0.8\textwidth]{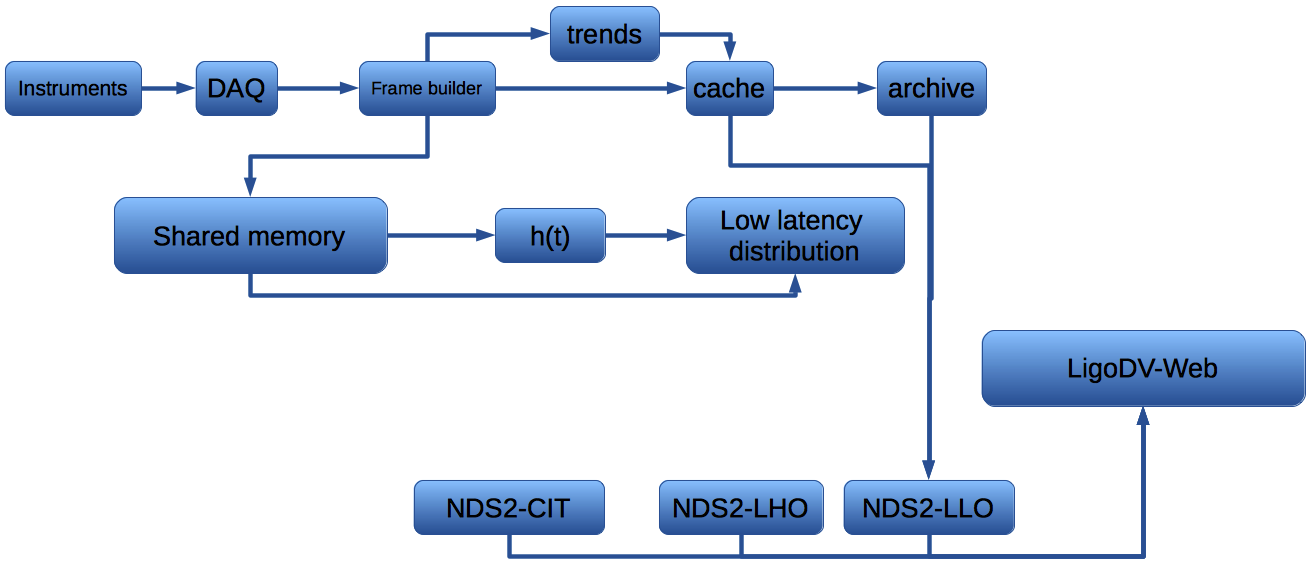}
  \caption{Data flow from instrument to the \ldvw\ application. The LIGO Hanford and Livingston Observatories collect data using local Data Acquisition Systems (DAQ), and write the data to a proprietary data storage format called Frame files. Low-latency data are written to shared memory and distributed to the local computing clusters. Frame data are cached to disk on the Tier 1 computing clusters. Both types of data are served to \ldvw\ upon request via the Network Data Servers (NDS2). 
    }
    \label{fig:dataflow}
\end{figure*}

Figure~\ref{fig:dataflow} diagrams the flow of data for the LIGO Hanford and LIGO Livingston Observatories.  The primary instrument at each observatory is a large modified Michelson laser interferometer with two 4\,km long arms that acts as a transducer from the stretching and squeezing of space-time caused by a passing gravitational wave (called strain because the stretching is proportional to the length, $h=\delta{}L/L$) into light power fluctuations at the output of the detector.  Each interferometer produces a calibrated time-series data stream known as $h(t)$ along with hundred of thousands of auxiliary data channels that record the behavior of the instrument and its environment. The Data Acquisition System (DAQ) collects analog and digital data with sample rates ranging from 1 to 32,000\,Hz, calculates some processed channels such as detector status channels, and passes them to the frame writer for immediate storage.  Post processing routines then create several other processed channels such as second and minute trends (max, min, mean, and root-mean-squared values per second and minute, respectively) as well as the calibrated $h(t)$ channel. With both Advanced LIGO observatories operating, the data acquisition rate is roughly 10MB/s or 1PB/year.

All of these data are written to local disk (cache system) in a proprietary format called Frame files~\cite{frames} while a subset of it is also distributed to a low-latency network.  Great care is needed to ensure data integrity and prevention of loss so the systems involved in data acquisition are tightly controlled and processes involved are limited to critical tasks. After the data are written to the local cache system several processes come into play.  Current data are kept in the cache system for 4 to 6 weeks, depending on available disk space, to allow rapid access for investigations and analysis. It is then is moved to an archive file system which is hierarchical in nature with a large tape archive and limited disk space.

The LIGO Data Replicator~\cite{ldr}, is responsible for distributing data to the many data centers and computing clusters involved in data analysis.  The Tier 1~\cite{tiern} data clusters are located at the Hanford and Livingston Observatories and LIGO Laboratory at Caltech.  They provide data for online data collection and analysis.  The observatories store data acquired on-site and some data from the other observatory used for analysis.  Caltech maintains all data with latencies varying from a few seconds for the low latency network to an hour or two for some trends. The LSC has other computer clusters designated Tier 2-4 that maintain storage of  data pertinent to current investigations.

Multiple Network Data Servers (NDS2) are available that provide access to data stored in frame files and in the low latency network at Tier 1 sites.  NDS2 provides critical back end services to the enterprise application (\ldvw).  NDS2 analyzes available frame files to produce channel lists and expected availability.  It responds to requests for channel lists, data availability and arbitrary intervals of data from specific channels. NDS2 uses a proprietary network protocol to communicate with clients over the Internet.  Users are authenticated using Kerberos against the LIGO online directory built on the Lightweight Directory Access Protocol (LDAP)~\cite{ldap}, which contains all collaboration members and approved client software packages such as \ldvw. Each NDS2 server provides access to frame files on disk and in the tape archive at one location.  A single location may have multiple NDS2 servers for increased throughput and reliability.  

\subsection{\ldvw\ overview}

 \ldvw\ is a web-based data viewer that provides access to data recorded at the LIGO Hanford, LIGO Livingston and GEO600 observatories, and the 40m prototype at Caltech. It grew out of the \ldv\ project (which itself grew out of DV6, developed for GEO600 by Martin Hewitson) which is a mature MATLAB program for viewing LIGO data with general analysis features. \ldv\ is supported on Macintosh, Windows and several Linux platforms but requires a relatively expensive Matlab license and installation of the NDS2 client.  These burdens, while not unreasonable, are enough to discourage casual investigation into the usefulness of \ldv. \ldvw\ was designed to greatly simplify access for all collaboration members by eliminating local installation and providing an interface that is compatible with standard internet devices and web browsers while providing many of the features of \ldv. \ldv is maintained as the more appropriate tool for in depth analysis.  \ldvw\ was thus designed to move all proprietary and commercial software to the server with the only requirements of the client being a modern web browser and Internet connection.  The browser requirements are common HTML5 features and Javascript support.  A side benefit is the availability of LIGO data on most Internet appliances including smartphones, tablets, Internet TV with browser capabilities as well as personal computers and workstations.


\section{Methods}
\label{Section:Methods}
\begin{figure*}
  \centering
  \includegraphics[width=0.6\textwidth]{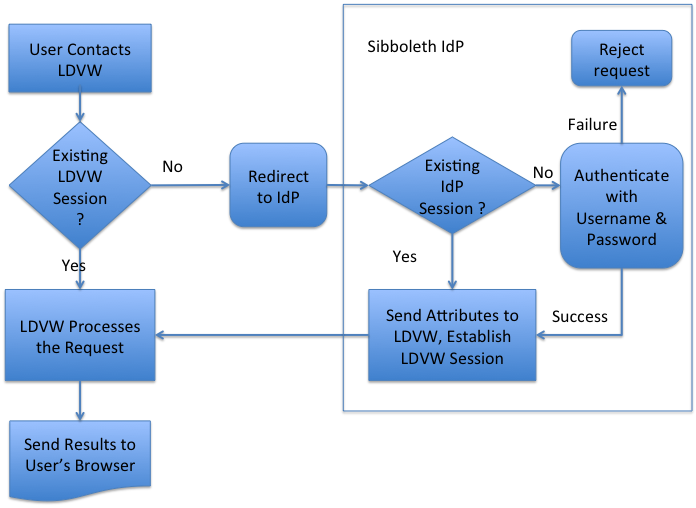}
  \caption{Flow of the Shibboleth Authentication Process. Shibboleth authenticates new sessions with a central Identity Provider to establish a new session. If the user has an existing session with any service provider there is no need to reenter credentials. If they do not have an existing session they are authenticated with username and password. 
    }
    \label{fig:shibflow}
\end{figure*}

\subsection{Authentication and authorization}

\ldvw\ is technically implemented as a Java Enterprise Application running in a Tomcat 7 container~\cite{tomcat}.  It uses Apache as a reverse proxy to leverage support for Shibboleth Single Sign On services allowing members to log into one supported service and have access to all other supported services for which they are authorized.  The back end relational database uses mySql to store local indexes of all available channels, a cache of recently transferred data, all plots created, and several log files and performance statistics.

Security is an important consideration for any web site.  The LSC deploys many web services with different access rules depending on the type of data presented.  \ldvw\ is only accessible to members of organizations in the collaboration that have a data sharing agreement with LIGO.

LIGO uses multiple technologies to implement secure Single Sign On (SSO) authentication and authorization:  Shibboleth, Grouper, Kerberos, LDAP, and the encrypted web protocol, https.  When combined with a custom program called myligo, this infrastructure provides a convenient, secure and powerful system to restrict access to the assets of the collaboration (for example, gravitational-wave data) without undue burden on the authorized users or the administrators.

A new person joining a group within the collaboration creates a ligo.org account on myligo, adding their contact information and group affiliation.  The principal investigator is then notified via email and prompted to log in to myligo to accept this person into their group. Once they have been accepted, they become an official group member, are added to the LSC roster.   

Shibboleth~\cite{morgan2004federated} is an open-source, standards-based system for SSO system authentication and authorization.  Shibboleth resources are broadly divided into two categories: Identity Providers (IdP) and Service Providers (SP).  A trust relationship is established between them through signed certificate based communication and sharing of certain metadata.  \ldvw\ is a Service Provider in this context. As such it knows nothing about its users but it trusts the LIGO Identity Providers.

Fig~\ref{fig:shibflow} shows the flow of the Shibboleth authentication process. When a user contacts a Service Provider, if they do not have an active session, they are redirected to the Identity Provider.  If they have a session with the Identity Provider they are redirected back to the Service Provider and the Identity Provider includes an appropriate set of attributes used to establish a user session with the Service Provider.  If no Identity Provider session is active, the user is prompted for a username and password, which is validated against their LDAP record.  Once a session is established with the Identity Provider, the user may establish a new session with any participating Service Provider without reconfirming their identity.  This provides the convenient Single Sign On process.  Subsequent requests to the Service Provider are processed immediately without further involvement of the Identity Provider.

Shibboleth also supports the use of multiple Identity Providers through a process known as Federation.  In that case each institution manages the identities of its members.  During the authentication process the user is redirected to the appropriate Identity Provider. \ldvw\ does not use this feature but may in the future if other gravitational wave research groups, such as KAGRA in Japan, enter data sharing agreements with LIGO.

The establishment of an active session gives the Service Provider a trusted identity of the user and additional attributes including a name, email address and group memberships. A user's group membership authorizes their access to certain parts of the \ldvw\ service.

Grouper~\cite{grouper} is a distributed access management system.  Privileges are granted or denied by membership in groups.  When members enter or leave the collaboration their identity is assigned to the appropriate groups. In the case of \ldvw\ the groups that are allowed access to the instrument data are determined by data sharing agreements.  The membership of those groups is managed by their Principal Investigators (PIs).  An additional group was created for \ldvw, the Service Provider, and subgroups were created by the developers to authorize certain users (developers and testers) for access to administrative functions and unreleased functions for testing.

Kerberos~\cite{kerberos} is an authentication system that is used in program-to-program communications by \ldvw\ but is not used directly by people accessing the web site.  The \ldvw\ application uses Kerberos to authenticate access to the Network Data Servers for data retrieval and programs that use the RESTful interfaces to \ldvw\ may use Kerberos to initiate their Shibboleth sessions without user interaction.  REST (REpresenttional State Transfer)~\cite{rest} is a technology that allows remote programs or scripts run in a browser to access services provided by \ldvw\ (e.g., to achieve automated production of plots for the time of a gravitational-wave event candidate).

\subsection{Network data services}

Frame files are designed to efficiently record large amounts of time series data from many channels in real time.  A typical file contains over 100,000 channels that have been directly acquired from the instrument or derived from acquired data.  The duration of a given frame file ranges between 1 second for data distributed for low latency analysis and 1 hour for minute trend data.  Trend data are generated from the raw time series with minute and second sample rates.  They record the minimum, mean, maximum, root-mean-squared and number of samples during each sample and are used to monitor the detectors' behavior over long periods.  Most frames containing raw data are 32 seconds long.  A channel is described by metadata that contains a name, sample rate, data type (integer, long, float, double, etc.) and may contain calibration factors and physical units.  Channels are sampled at rates that range between once per minute and 65,536\,Hz.

The efficiency of the frame file design is weighted heavily toward recording data at the expense of retrieval and analysis.  Recording the data is the real time process, while most retrieval and analysis are done without the same time constraints.  This storage scheme provides challenges for a service such as \ldvw, which presents data for a small number of channels over a long period of time as opposed to a large number of channels over a short period. The Network Data Server (NDS) was designed to ease such data access. 

There are two versions of the Network Data Server in operation, one is oriented toward real time data and is used in the control room of the observatories, known as NDS or NDS1.  NDS2 is mainly concerned with archived data although it does provide access to the low latency distribution network.  There is one NDS2 server at each observatory and one at the central archive maintained at Caltech.  A fourth NDS2 server is used by the  40\,m prototype interferometer at Caltech.  The NDS1 servers are unauthenticated and are only available on the local network of the observatories.  NDS2 servers use Kerberos backed by the collaboration's LDAP for secure access.

The NDS2 server and client software provides a more natural interface to the data in frame files stored on the local computer cluster.  Requests are made for a list of channels or data from a list of channels during a specific time interval.  The server maintains a list of channels, tracking changes over time.  When a request is made the appropriate frames are located and data for the requested channels is sent to the client.


\subsection{\ldvw}

\ldvw\ can be seen as another layer of abstraction on the NDS2 model; its primary purpose is to give users access to the large store of data in a natural, convenient manner without the need to understand how or where the data are stored. A user need only know which channel and what time to access the data. Thus it is not necessary for them to know how and where the data are stored.
\ldvw\ is typically used to help answer research questions about an interesting period of data, such as one containing a gravitational-wave candidate or poor quality data~\cite{detcharcompanion}. The purpose may be to identify the channels that are relevant to the question or to investigate the features present in those channels. 
Data are accessed by channel and time regardless of what type of frame contains the data and which servers host the frame files.  Data are presented in graphical form or as a downloadable file for use in local analyses. The design criteria for \ldvw\ are:
\begin{enumerate}
\item It presents a simple, easy to use interface to the user.
\item It requires only a modern browser (with common HTML5 components and javascript). It cannot require plugins or any special components to be loaded on the client side.
\item It provides robust, secure and fast access to as much gravitational wave data as possible.
\end{enumerate}

\begin{figure*}
  \centering
  \includegraphics[width=0.9\textwidth]{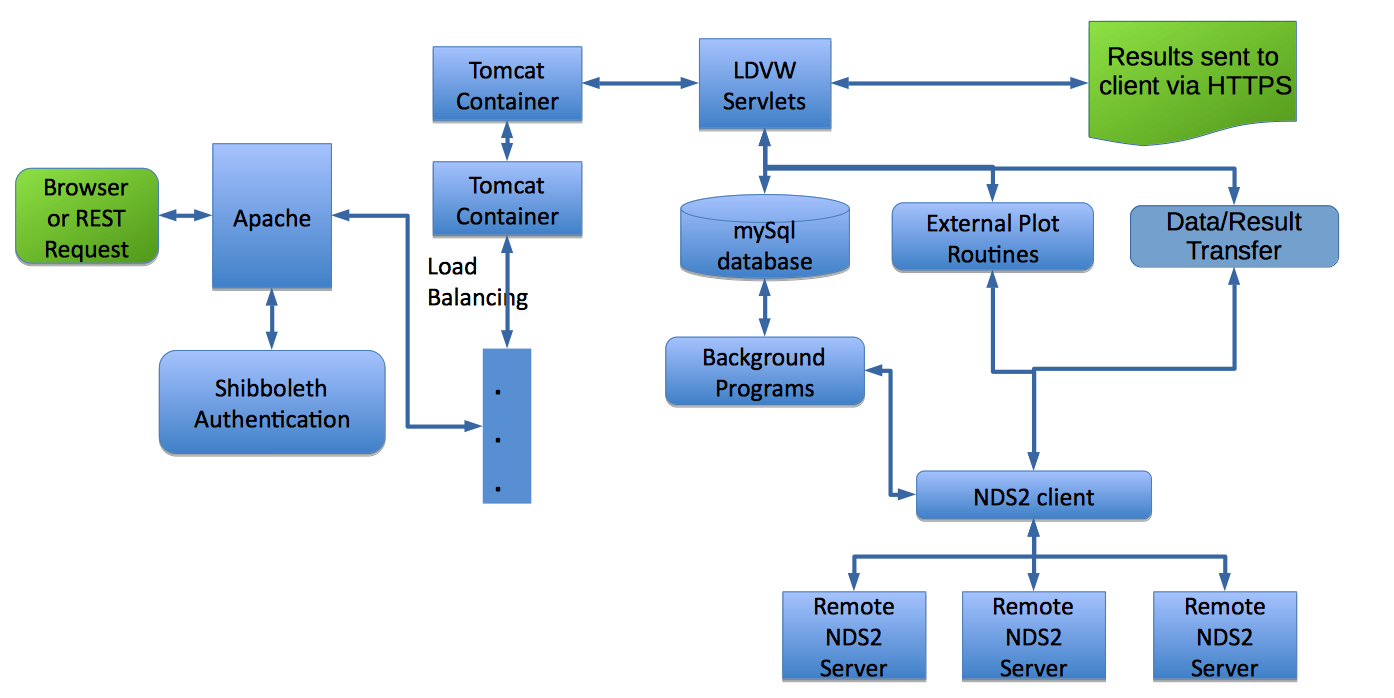}
  \caption{Overview of \ldvw\ Components. Apache is used to provide authentication and load balancing among multiple tomcat containers. The \ldvw\ servlets verify users are authorized, then handle the user requests by accessing the database and interfacing with external routines that produce plots or format data for transfers. The NDS clients retrieve the gravitational-wave data stored remotely.}
    \label{fig:ldvwflow}
\end{figure*}

\ldvw\ is implemented as Java Servlets (small programs that are run on the server to respond to user requests) that run inside the Tomcat containers.  The servlets handle the user interface, database interactions and some of the simpler plot products.  They also call external programs for the more complicated plot products.  There are also background tasks that do not interact directly with the user but facilitate timely interactions such as maintaining the channel database tables.

Figure~\ref{fig:ldvwflow} is a high-level block diagram of the \ldvw\ application.  Users make requests via a browser to a single Apache web server.  Apache handles the Shibboleth negotiations and only requests from authenticated users are passed to a Tomcat container which provide the Java Enterprise Application Environment for the \ldvw\ servlets.  The load-balancing feature of Apache allows easy expansion of capacity by adding more Tomcat containers to meet growing demand.  One Apache server is expected to handle as many Tomcat instances as needed for any foreseeable demand.

\begin{figure*}
  \centering
  \includegraphics[width=1.0\textwidth]{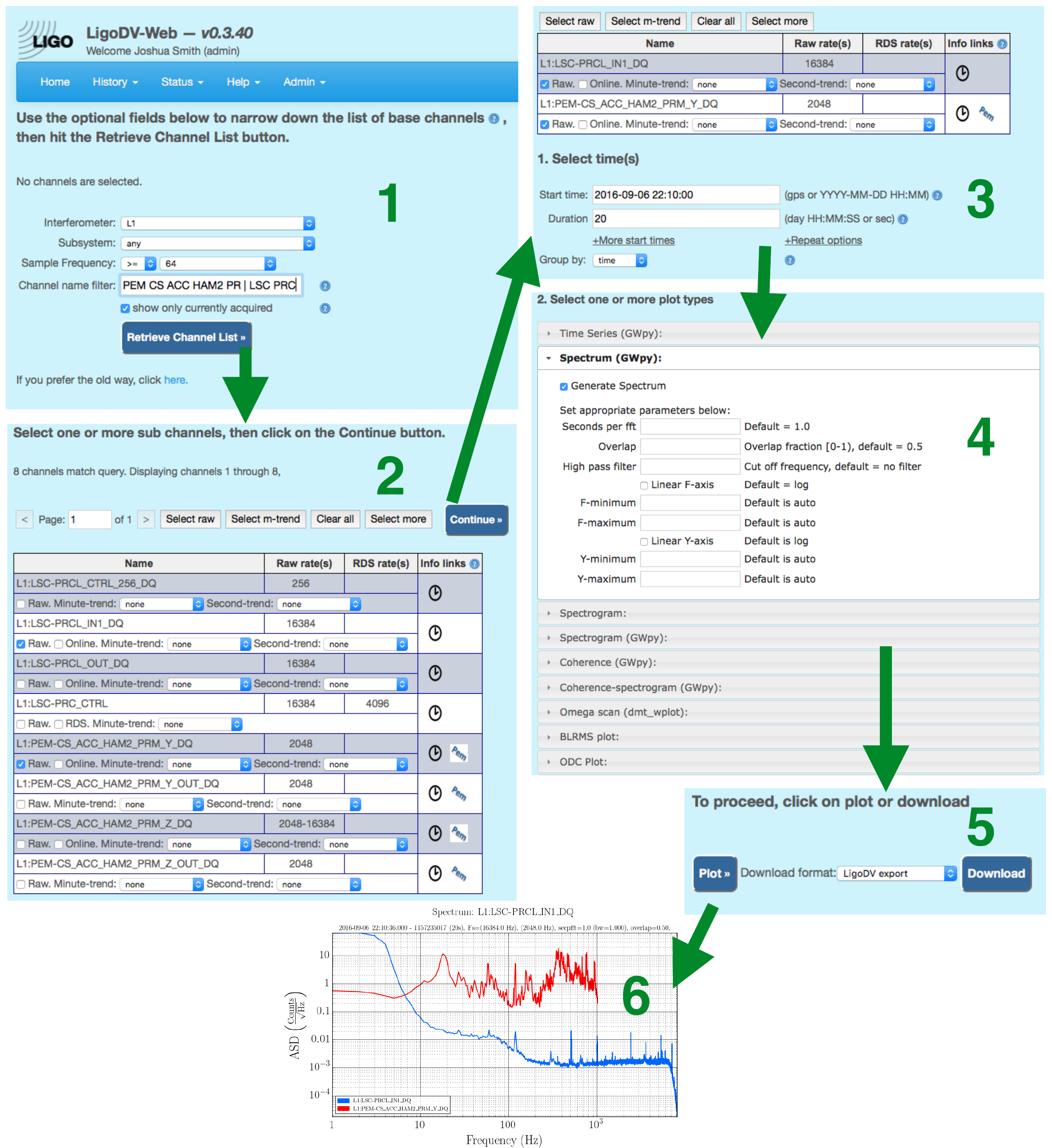}
  \caption{Example flow through the \ldvw\ user interface to generate a plot. 1) Users can enter one or more parameters to limit the search for channels. 2) Individual channels are then selected from the search results. 3) One or more time intervals are selected for the analysis as well as the way to group multiple channels and/or time intervals. 4) One or more plot products are chosen and the optional parameters are set. 5) Users can produce the plots specified or download raw data. 6) The resulting plots are presented as images that can be linked to or downloaded.
    }
    \label{fig:uiflow}
\end{figure*}

\subsection{Channels, database, and background processes}

\ldvw\ uses a mySql database to control most of the user interactions.  A major function of the database tables is to provide quick intuitive access to the data available via the NDS2 servers.  

The NDS2 servers provide lists of all available channels and are also capable of listing only those channels that have data available during a specific time period. Considering channels available from multiple servers, the current list contains over 14 million channels.  The complete list takes on the order of 10 minutes to transfer over university grade Internet connections.  However, NDS2 servers provide a hash of each channel type at each server, thus we only need to download channel lists when a change is detected.  

To simplify the channel list, \ldvw\ coalesces each channel into what is called a ``base channel." This represents the channel's raw data, 5 associated minute trends, 5 second trends, any reduced sample rate data and copies at multiple locations.  This consolidation reduces the effective channel list in \ldvw\ to about 811,000 base channels.  This greatly increases the speed of user searches for specific channels and makes it much easier to select the channels of interest needed to answer a specific research question.

\ldvw\ also keeps a semi-permanent database of all plots and analyses created.  It is less than permanent in the sense that the creator of a plot or an administrator can delete individual images but this is rarely done.  Each image has a permanent link that can be shared with others or used to include analyses in other web pages without downloading and copying the images themselves.  Images can also be grouped together under a user specified name (such as ``my favorite seismic spectrograms") and permanent links to these groups can also be shared or linked from other web resources. Links to images and image groups are a very popular means to share plots within the collaboration. 

The other background programs that assist this service perform functions such as monitoring the status of the web server and NDS2 servers, verifying the integrity of the database, and automatic backups of the tables in the database that cannot be recreated from the NDS2 channel lists and near real-time plot generation (described below).

\subsection{The User Interface}

The design requirements of the User Interface (UI) are that it be as easy to use as possible and require no software on the client side other than a modern browser that supports minimal HTML5 features and have JavaScript enabled.  This allows most Internet appliances to use \ldvw\ including smartphones, tablets as well as laptops and desktop workstations.  

The UI, shown in Figure~\ref{fig:uiflow}, provides the following major functions to non-administrative users: 
\begin{itemize}
\item Search the Channel database and select a subset of channels
\item Request one or more plot products (see Section~\ref{sec:plots}) for one or more channels at one or more times
\item Automatically upload images to the database and optionally view or group them
\item Review previous plots and plot groups made by any user
\item Check status of the NDS2 servers or channel data base
\item Provide links to information about specific channels and their availability 
\end{itemize}



\begin{figure*}
  \centering
  \includegraphics[width=1.0\textwidth]{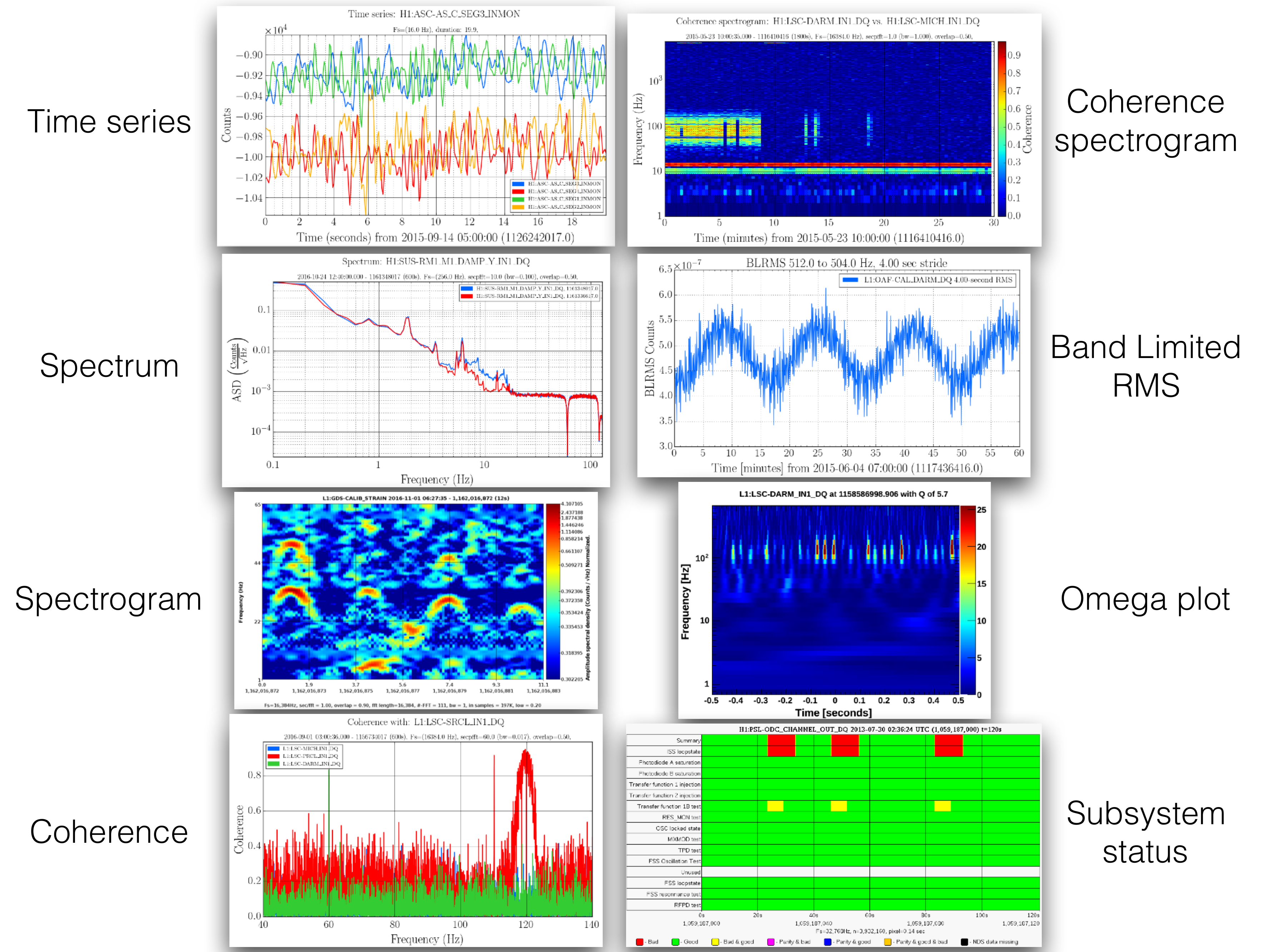}
  \caption{One example each, chosen from the database of user-made images, of the eight plot types available through \ldvw. 
    }
    \label{fig:ldvwplots}
\end{figure*}

\subsection{Plot products}\label{sec:plots}

The purpose of \ldvw\ is to provide quick easy access to gravitational-wave data not to provide comprehensive plotting and analysis tools.  Therefore we provide an expanding set of basic tools with the development focused on ease of operation and robustness. The reliability, ease of use and quality of plot products are of primary importance to our users.

Plotting is requested in one of three ways: a simple plot of one or more specific times, a recurring plot, or automatic near real-time displays.  Recurring plots use on-line data with a latency of one or two seconds.  Near real-time plots use a background job to create an self-updating web page that show status of the instrument a few minutes old, typically 5 minutes depending on the latency of raw frames being inserted into the NDS2 server’s available data at Caltech.  The main advantage of near-real time plots over recurring plots is they are served as static web pages, thus many people can view the same content simultaneously without overloading the server.

The current list of plot products available on \ldvw\ are listed below and shown in Figure~\ref{fig:ldvwplots}. The majority of these call on external programs to produce the plots. The use of external programs has increased robustness because they perform one function then exit. This means memory and other resource leaks are less damaging. Primary among these is the GWpy software package~\cite{gwpy}, an inter-connected software ecosystem that has significantly advanced GW detector characterization by enabling collaborators in the LSC to perform rigorous, complex analyses that were not previously possible, with minimal programming overhead. This package is at the heart of the LIGO Summary Pages~\cite{gwsumm} that have become the first port of call for a large fraction of the collaboration to access myriad statistics and figures or merit regarding operational sensitivity and performance of the LIGO and GEO600 detectors. Because the plots produced by this package are attractive and have benefited from feedback from many users, \ldvw\ has integrated them whenever possible. Additionally, \ldvw\ calls an external program to create ``Omega" plots, time-frequency representations of data based on Sine-Gaussian wavelets~\cite{Chatterji:2004qg}, which are also very widely used in the collaboration.

The currently available plot products are as follows. 
\begin{itemize} 
\item Time series (GWpy): Displays the time series on a linear-linear axis. Options include configurable $x-$ and $y-$axis limits, logarithmic $x-$ and $y-$axis scaling, and configurable high-pass filtering. 
\item Spectrum (GWpy): Displays an amplitude spectral density on a log-log axis. Options include seconds per Fast Fourier Transform (FFT), overlap fraction, high-pass filter and its cutoff frequency, $x-$ and $y-$axis limits and linear $x-$ and $y-$axis scaling.
\item Spectrogram: Displays a spectrogram (time-frequency representation with color as the $z-$ (amplitude/power) axis) on a linear-log axis. Options include windowing function type, scaling (amplitude, power, amplitude spectral density, power spectral density), seconds per FFT, overlap fraction, normalization, smoothing and interpolation, colormaps, $x-$, $y-$ and $z-$axis limits and linear $x-$ and $y-$axis scaling.
\item Coherence (GWpy): Displays the coherence versus frequency of two or more channels on a log-linear axis. Options include choice of reference channel (which channel the coherence of other channels will be computed with respect to), seconds per FFT, overlap fraction, high-pass filter and its cutoff, $x-$ and $y-$axis limits and linear $x-$ and $y-$axis scaling.
\item Coherence Spectrogram (GWpy): Displays a time-frequency representation with color representing coherence on the $z-$axis. Options are the same as for coherence, with the additional options of normalization and $z-$axis scaling. 
\item Omega plot (DMT Omega): Displays a whitened spectrogram centered at a user-specified time on a linear-log plot with color representing energy. Additional plots generated by the Omega algorithm can be optionally produced such as eventgrams and timeseries with no filtering, whitening, or autoscaling. Options include plot time ranges, downsampling frequency, search time, frequency and Q range, and $y-$ and $z-$axis limits.
\item Band Limited RMS (BLRMS) (GWpy): Displays a timeseries of BLRMS on a linear-linear axis. Options include upper and lower cutoff frequencies for the bandpass filter used, and the duration of each RMS calculation. 
\item Detector or Subsystem Status: Displays a colored bar bitmap timeseries of the status of the detector or a given detector subsystem, with colors indicating a good, bad or unknown status at a given time.
\end{itemize} 

Timeseries and spectrum data are downloadable as comma separated values files or as MATLAB structures. Additionally, timeseries data can be converted into audio files and downloaded. Expanding data download capabilities to more of the above plot types is currently a high priority. 


\section{Status and future}
\label{Section:Conclusions}

\begin{figure*}
  \centering
  \includegraphics[width=0.8\textwidth]{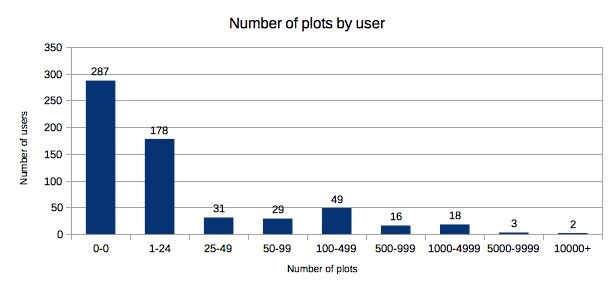}
  \caption{The number of people (unique users) that have made a given number of plots on \ldvw. Of the more than 1000 LSC members 613 members have used \ldvw. Of those 287 have not created plots of LIGO data but only visited the site likely to view links to plots made by others. Of the 326 users that have made plots about half have made less than 25 plots. Many people use the site regularly: 88 people have produced more than 100 plots each and the top users have created more than 10,000 plots each. 
    }
    \label{fig:ldvwplotsbyuser}
\end{figure*}

\begin{figure*}
  \centering
  \includegraphics[width=0.9\textwidth]{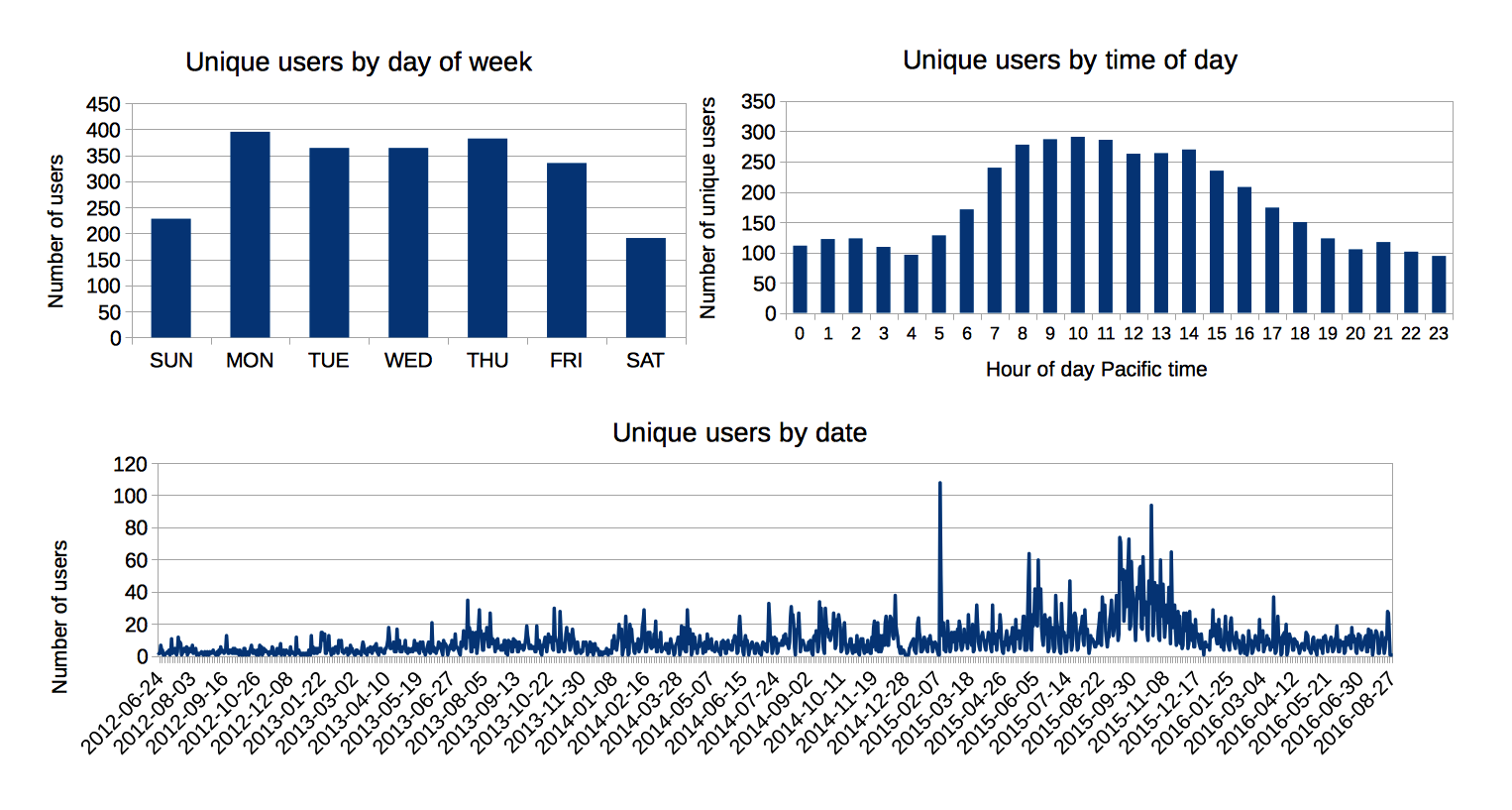}
  \caption{Usage patterns of \ldvw. During a typical visit to the website a user may generate a plot, check channel availability, download data, or view plots made by others. All graphs show the number of people (unique users) who visit the site for that period. \textit{Upper left:} by day of the week; \textit{Upper right:} by hour of the day; \textit{Lower:} day by day since the launch of the website. The broad bump in daily users in September 2015 coincides with the first direct detection of gravitational waves. The LSC has more than 1000 members, so these plots give an indication of the acceptance of \ldvw.
  } 
    \label{fig:ldvwusage}
\end{figure*}
  
\ldvw\ is a successful project that has provided easy, secure and universal access to a large distributed scientific data store for the LIGO Scientific Collaboration. Since being made available to the LSC in 2012, \ldvwusers\ LSC members have logged into \ldvw\ in \ldvwsessions\ sessions and have produced a total of \ldvwplots\ plots, with the ten most heavy users having generated more than 2,000 plots each, as shown in Figure~\ref{fig:ldvwplotsbyuser}. A snapshot of the statistics for when people typically use \ldvw\ is shown in Figure~\ref{fig:ldvwusage}. The usage varies greatly depending on how interesting recent data are. Around the discovery of GW150914 the website was visited by 40-90 unique users per day, while the mean number of unique users per day is around 10. These usage patterns imply acceptance by the community.

To build upon this, future development of \ldvw\ will be focused on expanding the number of plot products and making their options and presentation more consistent by leveraging GWpy and to make data downloadable for all plot products. Development will be done to add user preferences, such as favorite channel lists and settings. Additionally, load balancing will be expanded to handle the growing number of users and in anticipation of future public access to the LIGO data store. 

\section*{Acknowledgments}
We thank our colleagues in the LIGO Scientific Collaboration and Virgo Collaboration for providing many fruitful ideas, bug reports, and feature requests for \ldvw\ and for thoughtful input on this manuscript. Thanks to TJ Massinger for careful review of the manuscript leading to many improvements. This work was supported by National Science Foundation awards PHY-1104371 and PHY-0600953. 

\bibliographystyle{unsrt}
\bibliography{references}

\end{document}